\documentclass{article}
\usepackage{amsmath}

\DeclareMathOperator{\Tr}{Tr}
\newcommand{\e}{\,\mathrm{e}}
\newcommand{\pdc}[3]{\ensuremath{\left(\frac{\partial #1}{\partial #2}\right)_{#3}}}
\newcommand{\avrg}[1]{\ensuremath{\left<#1\right>}}
\newcommand{\avrgo}[1]{\ensuremath{\left<#1\right>_{1,2}}}
\newcommand{\const}{\mathrm{const}\,}

\newcommand{\vb}{\ensuremath{\mathbf{v}}}
\newcommand{\rs}{\ensuremath{\rho_\text{s}}}

\title{Supersolidity of glasses}
\author{A.\,F.\,Andreev\footnote{e-mail: andreev@kapitza.ras.ru}}
\date{}

\begin{document}
\maketitle

\begin{center}
Kapitza Institute for Physical Problems, Russian Academy of Sciences,\\
Kosygina 2, Moscow, 119334 Russia
\end{center}

\begin{abstract}Supersolidity of glasses is explained as a property of an
unusual state of condensed matter. This state is
essentially different from both normal and superfluid solid
states. The mechanism of the phenomenon is the transfer of mass by
tunneling two level systems.

PACS: 67.80.-s, 67.80.Mg, 67.40.Kh
\end{abstract}

\section{Introduction}
It was shown theoretically \cite{lit1,lit2,lit3} that owing to the large
probability of quantum tunneling of atoms, solid helium may be
superfluid. All attempts to observe the superflow experimentally were
unsuccessful (see \cite{lit4} and \cite{lit5}).

Kim and Chan \cite{lit6} observed the reduction
of solid $^4$He rotational inertia
below $0.2K$ in the torsional oscillator
experiments and interpreted it as the superfluidity of the solid.
Further experiments (see \cite{lit7} and references therein) show
that the superfluid fraction observed for highly disordered
(glassy) samples is remarkably large, exceeding $20\%$. This
fraction seems to be absent in ideal helium crystals.

In 1972 it was shown \cite{lit8,lit9} that the quantum
tunneling of the atoms explains some low
temperature properties (thermal, electromagnetic, and acoustic) of
glasses. The key point is the presence of the so-called tunneling two
level systems (TLS) in the solid. A TLS can be understood as an
atom, or a group of atoms, which can tunnel between two localized
states characterized by a small energy difference.

In this paper we show that owing to the presence of coherent TLS's,
quantum glasses manifest peculiar properties which are essentially
different from those of normal and superfluid solids.  Precisely these
peculiar properties are observed experimentally (both \cite{lit4,lit5}
and \cite{lit6,lit7}). At present the terms ``supersolid'' and
``supersolidity'' are used simply as synonyms for ``superfluid solid''
and ``superfluidity of solids'' respectively (see \cite{lit10} for a
review). We propose to use the term ``supersolidity'' to refer to the
above mentioned properties of quantum glasses.

A normal solid is characterized by a single velocity of
macroscopic motion: the solid bulk velocity \vb. The
momentum density is $\rho \vb $, where $\rho$ is the mass
density. The general motion of a superfluid solid is characterized
(see \cite{lit1}) by two mutually independent velocities: that
of the solid bulk and the superfluid one. The supersolid
(in our sense of the word) is characterized by a single velocity
\vb\ of the solid bulk, but under certain conditions (see
below) the momentum density is $(\rho - \rs )\vb$, where
$\rs /\rho$ is the supersolid fraction. This is exactly what we
need to explain both the reduction of rotational inertia
\cite{lit6,lit7} and the absence of a superflow
\cite{lit4,lit5}. We calculate \rs\ in terms of TLS
parameters. The supersolid fraction, being proportional to the
squared TLS tunneling amplitude, can be considerable for
highly disordered solid $^4$He and other quantum solids (hydrogen).

Our results are supported in recent experiment by Grigorev et al.
\cite{lit11}. They measured the temperature dependence of pressure
in solid $^4$He grown by the capillary blocking technique. At
temperatures below $0.3K$ (where the supersolidity was observed) they
found the glassy $\propto T^2$ contribution to pressure.
This is exactly what one expects from the TLS.

\section{TLS in moving glasses}
The Hamiltonian $H_0$ of a given TLS in the frame of reference in
which the solid bulk velocity \vb\ is zero, can be written
as
\begin{equation*}
H_0 = -\varepsilon \sigma_3 + J\sigma_1.
\end{equation*}
Here $\mp\varepsilon$ ($\varepsilon > 0$) are energies of two
localized states, $J$ is the tunneling amplitude, and $\sigma_i$
($i=1,2,3)$ are the Pauli matrices.

Let us suppose that the tunneling of the TLS is accompanied by
displacement of a mass $m$ by a vector $\mathbf{a}$. The
coordinates $\mathbf{r}_{1,2}$ of the center of gravity of the TLS
before and after the tunneling can be written as $\mathbf{r}_{1,2}
= \mp \mathbf{a}/2$. The operator form of the last equality is
$\mathbf{r} = -\sigma_3 \mathbf{a}/2 $. The operator of velocity
is determined by the commutator:
\begin{equation*}
\dot{\mathbf{r}} =
  \frac{i}{\hbar}[H_0,\mathbf{r}] =
 -\frac{J\mathbf{a}}{\hbar}\sigma_2.
\end{equation*}
The TLS momentum in the frame in which $\vb  = 0$, is
\begin{equation*}
\mathbf{p} = 
 m\dot{\mathbf{r}} =
-\frac{mJ\mathbf{a}}{\hbar}\sigma_2.
\end{equation*}
In an arbitrary frame of reference a description  of the TLS by
means of a discrete coordinate is impossible. But we can use
Galilean transformations to find the TLS Hamiltonian and momentum
in the frame in which \vb\ is finite. We obtain
\begin{align*}
H_0 &\rightarrow H_0 + \mathbf{p}\vb  + mv^2/2,\\
\mathbf{p} &\rightarrow \mathbf{p} + m\vb ,
\end{align*}
respectively. The last terms of both expressions must be included
to the total kinetic energy and momentum of the solid bulk.
Therefore, the contributions of the TLS tunneling to the energy
and momentum of entire system are
\begin{equation}
H = H_0 + \mathbf{p}\vb
\label{eq5}
\end{equation}
and $\mathbf{p}$, respectively.
These two operators represent the energy and momentum of the
tunneling TLS in the solid moving with velocity \vb.
Note that the operators $\mathbf{p}$ and $H$ do not commute with
each other.

The eigenvalues of the Hamiltonian $H$ are $E_{1,2} = \mp E$,
where $E = \left(\varepsilon^2 +\Delta^2\right)^{1/2}$,
$\Delta = J\left(1+u^2\right)^{1/2}$, and 
$u = (m/\hbar)(\mathbf{a}\vb)$.
According to the general result of quantum mechanics
(\cite{lit12}, \S11) the mean values of momentum
\avrgo{\mathbf{p}} in the stationary states 1 and 2 are
\begin{equation*}
\avrgo{\mathbf{p}} = \avrgo{\frac{\partial H}{\partial \vb }}
= \frac{\partial E_{1,2}}{\partial \vb }.
\end{equation*}
We have
\begin{equation*}
\avrgo{\mathbf{p}} = \mp \frac{J^2 m^2}{\hbar^2
E}\mathbf{a}(\mathbf{a}\vb ).
\end{equation*}
In the case of nonzero \vb, the TLS has nonzero mean
values of momenta in both stationary states. Note that
in the TLS ground state, the projection of the momentum
$\avrg{\mathbf{p}}_1$ on the direction of velocity \vb\ is
negative. This is the mechanism of supersolidity.
The Hamiltonian $H$ is identical to that of spin $1/2$
in magnetic field. The sign of $\avrg{\mathbf{p}}_1$
corresponds to Pauli paramagnetism.

\section{Supersolidity}
The equilibrium density matrix of the TLS (which is an almost
closed system) in a uniformly rotating frame is
\begin{equation*}
\e^{(f'-H')/T},
\end{equation*}
where $f'$ and $H'$ are the free energy and Hamiltonian in this
frame. The latter is determined by the expression
\begin{equation*}
H' = H - \mathbf{\omega} \mathbf{M} = H_0 + \mathbf{p} \vb 
- \mathbf{\omega} \mathbf{M},
\end{equation*}
where $\mathbf{\omega}$ is the angular velocity, $\mathbf{M}
=\mathbf{R} \times \mathbf{p}$ is the TLS angular momentum,
$\vb  = \mathbf{\omega} \times \mathbf{R}$, and $\mathbf{R}$
is the TLS coordinate with respect to the rotation axis. We obtain
$H' = H_0$. This means that a uniformly rotating supersolid behaves
like a normal solid.

However, suppose that the solid bulk velocity depends on time
$\vb =\vb (t)$ and is ``switched on'' adiabatically.
This means (see \cite{lit13}, \S 11) that the switching time is
much longer than the relaxation time in the solid but much shorter
than the time during which the solid can be regarded as thermally
insulated. The second of these two characteristic times is very
long due to the Kapitza thermal resistance.

According to the general result of statistical mechanics
(\cite{lit13}, \S11 and \S15) we have
\begin{equation*}
\avrg{\mathbf{p}} = \avrg{\frac{\partial H}{\partial \vb }} =
\pdc{f}{\vb}{T},
\end{equation*}
where
\begin{equation}
f = -T\log \left(\Tr \e^{-H/T}\right)
\label{eq11}
\end{equation}
is the TLS free energy and $H$ is determined by \eqref{eq5} with
$\vb=\vb (t)$.

The free energy \eqref{eq11} can be written as
\begin{equation*}
f = -T\log\left(\e^{-E_1/T} + \e^{-E_2/T}\right).
\end{equation*}
Here $E_{1,2} = \mp E$ are the eigenvalues of the Hamiltonian $H$.
The mean value of the TLS momentum is
\begin{equation*}
\avrg{\mathbf{p}} = \frac{m\mathbf{a}}{\hbar}
\pdc{f}{u}{T}.
\end{equation*}
Simple calculation gives
\begin{equation*}
\pdc{f}{u}{T} = - \frac{J^2 u}{E}\tanh{\frac{E}{T}}
\end{equation*}
or
\begin{equation*}
\avrg{p_i} = - m_{ik}^{(s)} v_k,
\end{equation*}
where
\begin{equation*}
m_{ik}^{(s)} = \left(\frac{Jm}{\hbar}\right)^2 a_i a_k
\frac{\tanh(E/T)}{E}.
\end{equation*}

Let $Nd\varepsilon$ ($N = \const$) is the number of TLS's per unit
volume of the solid and per interval of the energy half-difference
$d\varepsilon$ near some $\varepsilon$ which is much smaller than
the characteristic height $U$ of the energy barriers in the solid.
The total momentum density $\mathbf{j}$ is
\begin{equation*}
j_i = \rho v_i - \rho_{ik}^{(s)} v_k,
\end{equation*}
where the supersolid density tensor is
\begin{equation*}
\rho_{ik}^{(s)} = \avrg{m^2 J^2 a_i a_k}\frac{N}{\hbar^2}
\int\limits_{\max(\Delta,T)}^{U}\frac{d\varepsilon}{\varepsilon}.
\end{equation*}
Here \avrg{\dots} means the averaging over the ensemble of TLS's, and
$\max(\Delta, T)$ is of the order of $\Delta$ if $T\ll\Delta$ and
of the order of $T$ if $T\gg\Delta$. Both $T$ and $\Delta$ are
much smaller than $U$.

For an isotropic system (glass) we have $\rho_{ik}^{(s)} = \rs 
\delta_{ik}$, where
\begin{equation*}
\rs = \frac{N}{3 \hbar^2} \avrg{m^2 J^2 a^2}
\log\frac{U}{\max(\Delta,T)}.
\end{equation*}

We see that the characteristic temperature of supersolidity is of
the order of $\Delta$. The critical velocity $v_c$ is determined
by the condition $u_c \sim 1$. We have $v_c \sim \hbar/(ma)$.
The critical velocities observed experimentally \cite{lit6} are
very small. This suggests the macroscopic character of the most
effective TLS's. In principle, this is possible. The pressure
dependence of \rs\ is determined by the competition of all
parameters $N$, $m$, $J$, and $a$. 
Efficient TLS tunneling is facilitated by
the presence of a region with reduced local
density in the vicinity of the TLS.
The $^3$He impurity, due to the
smaller mass of $^3$He atoms, must bind to such regions (see
\cite{lit10}) destroying TLS. This is a simple explanation of
supersolidity suppression by $^3$He impurities observed in the
experiments \cite{lit6}.

\section{Acknowledgments}
I thank L.A.Melnikovsky for helpful discussions. This work was
supported by the Russian Foundation for Basic Research, project
no. 06-02-17369a and by grant NSh-7018.2006.2 under the Program
for Support of Leading Science Schools.

\end {document}